\documentclass{aipproc}
\layoutstyle{6x9}
\newcommand{\gen}{$G_{E}^{n}~$}
\newcommand{\gmn}{$G_{M}^{n}~$}
\begin{document}
\title{Measurement of the Charge Form Factor of the Neutron \gen from ${\vec{d}(\vec{e},e'n)p}$ at $Q^{2}=0.5$~and~$1.0~(GeV/c)^{2}$ } 

\author{N. Savvinov, for the E93-026 JLab collaboration}{
  address={Department of Physics, University of Maryland, College Park, USA}
}

\date{\today}

\begin{abstract}
We determined the electric form factor of the neutron
$G_E^n$ via the reaction ${\vec{d}(\vec{e},e'n)p}$ using a
longitudinally polarized electron beam and a frozen 
polarized $^{15}ND_3$ target at Jefferson Lab.
The knocked out neutrons were detected in a segmented
plastic scintillator in coincidence with the
quasi-elastically scattered electrons which were tracked in
Hall C's High Momentum Spectrometer.
To extract $G_E^n$, we compared the experimental
beam--target asymmetry with theoretical calculations based
on different $G_E^n$ models.
We report the preliminary results of the fall 2001 run at
$Q^{2}~=~0.5~$~and~$1.0~(GeV/c)^{2}$.
\end{abstract}

\maketitle

\section{Introduction}

In a non-relativistic picture, the charge form-factor of the neutron \gen is related to the charge distribution in the neutron and thus is important for our understanding of electromagnetic structure of nucleons. Despite a great effort focused on its determination, \gen remains poorly known.  Two major difficulties faced by experimenters in their studies of \gen are its small magnitude and the lack of a free neutron target. Unpolarized cross-section measurements, from which \gen was extracted until 1990's, were incapable of overcoming these difficulties and yielded inconsistent or model-dependent results.
 
Recent technological advances in high duty factor accelerators, polarized sources, polarized targets and recoil polarimetry made possible  double polarization methods of determining $G_{E}^{n}$. These methods use asymmetries rather than cross-sections and thus reduce sensitivity to systematic errors. 

In the experiment described here a longitudinally polarized electron beam was scattered off a polarized deuterated ammonia target. The polarized electron-neutron scattering cross section consists of helicity-independent and helicity-flip terms. The helicity induced asymmetry, given by the ratio of these two terms, depends on \gen. For a general orientation of the target polarization this dependence is complex.  In order to minimize the influence of the magnetic form factor \gmn and maximize the sensitivity to \gen, the direction of the target polarization was chosen to be perpendicular to the three-momentum transfer $\vec{q}$ and to lie in the scattering plane. For this case the expression for the electron-neutron asymmetry $A_{en}$ simplifies to the following:

$$ A_{en} = \frac{-2\sqrt{\tau(1+\tau)}G^{n}_{M}G^{n}_{E}}{(G^{n}_{E})^{2}+\tau(1+2(1+\tau)\tan^{2}(\theta_{e}/2))(G^{n}_{M})^{2}}\;\;,$$

\noindent where  $\tau = \frac{Q^{2}}{4M^{2}}$, $M$ is the neutron mass, and $\theta_{e}$ is the electron scattering angle.

Since neutrons in deuterium are not free, the measured electron-deuteron asymmetry $A^{V}_{ed}$ differs from $A_{en}$ due to reaction mechanisms such as final state interactions and meson exchange currents. These reaction mechanisms were taken into account in theoretical calculations used for extraction of \gen \cite{arenh}.

\section{Experimental setup}

The experiment E93-026 was conducted in Hall C of Thomas Jefferson Accelerator Facility (Jefferson Lab) in 1998 and 2001.  The measurements were taken at two points,  $Q^{2}=0.5$ and $1.0~(GeV/c)^{2}$. 

The beam polarization was measured using a Moeller polarimeter. The average beam polarization during the experiment was 75\%. The beam current was limited to $100~nA$ to avoid excessive thermal and radiation damage to the target polarization. A system of raster magnets was used to distribute these stresses uniformly over the full target cell. The readings of the raster magnets were also used by the reconstruction algorithm to determine the horizontal and vertical position of the interaction point. 

The solid polarized target \cite{crabb} was developed by University of Virginia and was successfully used in two experiments prior to being used in E93-026. The basic components of the polarized target include a superconducting magnet operated at 5 Tesla, a $^{4}He$ evaporation refrigerator, a pumping system, a high power microwave tube operating at frequencies around 140 GHz and an NMR system for measuring the target polarization. The target material was polarized using the principle of dynamic nuclear polarization. Target polarization was determined by measuring the impedance change of the series resonant LCR circuit due to the nuclear magnetic moment. The conversion constants between the area of the NMR signal and the target polarization were obtained by a series of thermal equilibrium measurements. The target polarization typically varied between 15\% and 35\% and averaged to 22\%. 

After interaction in the target material, the scattered electrons were detected in the High Momentum Spectrometer of Hall C. Recoil nucleons were detected in the neutron detector which consisted of six planes of thick scintillators and two planes of thin scintillators. The latter were used for particle identification. The detector was set along the direction of the three-momentum transfer and was enclosed in a concrete hut open towards the target. The neutron vertical position was determined by the segmentation of the detector while the horizontal position was determined from the time difference of the phototubes. 

\section{Analysis and results}

\begin{figure}
\centering
\includegraphics[scale=0.8]{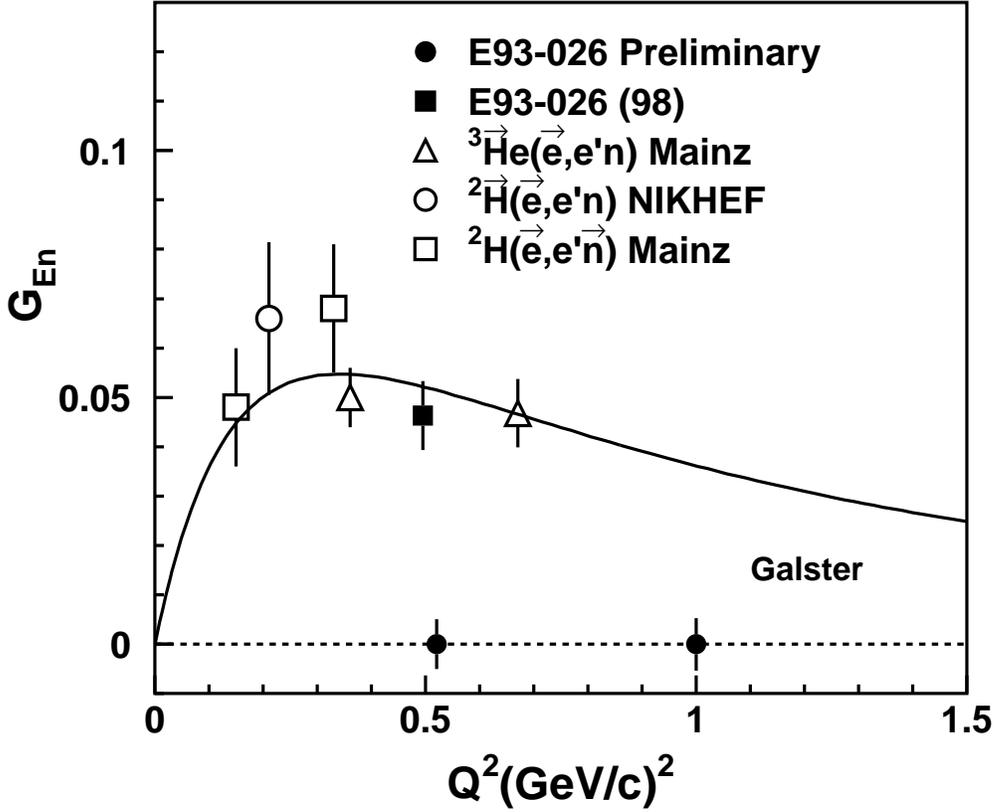}
\caption{Double polarization world data on \gen. Filled circle: preliminary E93-026 2001 data (error bars only), filled square: final E93-026 result for 1998 data \cite{zhuhg}, open triangles: polarized helium data \cite{golak}, open squares: recoil polarimetry data \cite{herberg,ostrick}, open circle: NIKHEF polarized deuterium data \cite{passchier}.}
\end{figure}

The electrons in the HMS were reconstructed using the standard HMS reconstruction code extended for the effects of beam raster and target magnetic field. On the neutron detector side a custom tracking algorithm was developed for proper particle identification. 
Neutrons were defined as events with no hits in the paddles along the track to the target, within a narrow time interval and within a 100 MeV range of invariant mass around the nucleon mass. A number of other cuts was applied to optimize the dilution factor and limit the recoil momentum to values where nuclear corrections are small. 

After event reconstruction and event selection, charge and dead-time normalized yields were produced for each beam helicity state, $N^{+}$ and $N^{-}$. From these yields the raw asymmetry $\epsilon$ was calculated:

$$\epsilon=\frac{N^{+}-N^{-}}{N^{+}+N^{-}}=P_{B}P_{T}f A^{V}_{ed}.$$

In order to obtain $A^{V}_{ed}$ from the raw asymmetry $\epsilon$, one needs the knowledge of dilution factor $f$ (due to scattering from unpolarized materials) in addition to beam and target polarizations $P_{B}$ and $P_{t}$. Dilution factor was calculated using montecarlo simulations. Finally, $A^{V}_{ed}$ was corrected for accidental background and radiative effects.

The \gen was extracted by comparing the corrected experimental asymmetry to the theoretical asymmetry averaged over the experimental acceptance under different assumptions about the size of the \gen.  Preliminary results are consistent with the Galster parametrization. The systematic error is expected to be dominated by uncertainty in target polarization (3-5\%) and dilution factor ($\sim3\%$).

\bibliographystyle{aipproc}
\bibliography{paper}

\begin{thebibliography}{7}
\expandafter\ifx\csname natexlab\endcsname\relax\def\natexlab#1{#1}\fi
\providecommand{\enquote}[1]{``#1''}
\expandafter\ifx\csname url\endcsname\relax
  \def\url#1{\texttt{#1}}\fi
\expandafter\ifx\csname urlprefix\endcsname\relax\def\urlprefix{URL }\fi

\bibitem[Arenh\"{o}vel et~al.(1988)]{arenh}
Arenh\"{o}vel, H., et~al., \emph{Z. Phys.}, \textbf{A331}, 123 (1988).

\bibitem[Crabb et~al.(1995)]{crabb}
Crabb, D., et~al., \emph{Nucl. Instr. Meth.}, \textbf{A356}, 9 (1995).

\bibitem[Zhu et~al.(2001)]{zhuhg}
Zhu, H., et~al., \emph{Phys. Rev. Lett.}, \textbf{87}, 081801 (2001).

\bibitem[Golak et~al.(2001)]{golak}
Golak, J., et~al., \emph{Phys. Rev.}, \textbf{C63}, 034006 (2001).

\bibitem[Herberg et~al.(1999)]{herberg}
Herberg, C., et~al., \emph{Eur. Phys. J.}, \textbf{A5}, 131 (1999).

\bibitem[Ostrick et~al.(1999)]{ostrick}
Ostrick, M., et~al., \emph{Phys. Rev. Lett.}, \textbf{83}, 276 (1999).

\bibitem[Passchier et~al.(1999)]{passchier}
Passchier, I., et~al., \emph{Phys. Rev. Lett.}, \textbf{82}, 4988 (1999).

\end{thebibliography}

\end{document}